\font\ottorm=cmr8

\newcount\mgnf\newcount\tipi\newcount\tipoformule
\newcount\aux\newcount\piepagina\newcount\xdata
\mgnf=0
\aux=0           
\tipoformule=1   
\piepagina=2     
\xdata=0         
\def\Di{21\ ottobre\ 1997}

\ifnum\mgnf=1 \aux=0 \tipoformule =1 \piepagina=1 \xdata=1\fi
\newcount\bibl
\ifnum\mgnf=0\bibl=0\else\bibl=1\fi
\bibl=0

%
%
%
%
\ifnum\bibl=0
\def\ref#1#2#3{[#1#2]\write8{#1@#2}}
\def\rif#1#2#3#4{\item{[#1#2]} #3}
\fi

\ifnum\bibl=1
\openout8=ref.b
\def\ref#1#2#3{[#3]\write8{#1@#2}}
\def\rif#1#2#3#4{}

\fi

\def\9#1{\ifnum\aux=1#1\else\relax\fi}
\ifnum\piepagina=0 \footline={\rlap{\hbox{\copy200}\
$\st[\number\pageno]$}\hss\tenrm \foglio\hss}\fi \ifnum\piepagina=1
\footline={\rlap{\hbox{\copy200}} \hss\tenrm \folio\hss}\fi
\ifnum\piepagina=2\footline{\hss\tenrm\folio\hss}\fi

\ifnum\mgnf=0 \magnification=\magstep0
\hsize=13.5truecm\vsize=22.5truecm \parindent=4.pt\fi
\ifnum\mgnf=1 \magnification=\magstep1
\hsize=16.0truecm\vsize=22.5truecm\baselineskip14pt\vglue5.0truecm
\overfullrule=0pt \parindent=4.pt\fi

\let\a=\alpha\let\b=\beta  \let\d=\delta
\let\e=\varepsilon \let\z=\zeta \let\h=\eta
\let\k=\kappa \let\l=\lambda  \let\n=\nu
\let\x=\xi \let\p=\pi \let\r=\rho  \let\t=\tau
 \let\f=\varphi  \let\o=\omega
 \let\G=\Gamma  
   \let\F=\Phi
  
{\count255=\time\divide\count255 by 60 \xdef\oramin{\number\count255}
\multiply\count255 by-60\advance\count255 by\time
\xdef\oramin{\oramin:\ifnum\count255<10 0\fi\the\count255}}
\def\ora{\oramin }

\ifnum\xdata=0
\def\data{\number\day/\ifcase\month\or gennaio \or
febbraio \or marzo \or aprile \or maggio \or giugno \or luglio \or
agosto \or settembre \or ottobre \or novembre \or dicembre
\fi/\number\year;\ \ora}
\else
\def\data{\Di}
\fi

\setbox200\hbox{$\scriptscriptstyle \data $}
\newcount\pgn \pgn=1
\def\foglio{\number\numsec:\number\pgn
\global\advance\pgn by 1} \def\foglioa{A\number\numsec:\number\pgn
\global\advance\pgn by 1}
\global\newcount\numsec\global\newcount\numfor \global\newcount\numfig
\gdef\profonditastruttura{\dp\strutbox}
\def\senondefinito#1{\expandafter\ifx\csname#1\endcsname\relax}
\def\SIA #1,#2,#3 {\senondefinito{#1#2} \expandafter\xdef\csname
#1#2\endcsname{#3} \else \write16{???? ma #1,#2 e' gia' stato definito
!!!!} \fi} \def\etichetta(#1){(\veroparagrafo.\veraformula) \SIA
e,#1,(\veroparagrafo.\veraformula) \global\advance\numfor by 1
\9{\write15{\string\FU (#1){\equ(#1)}}} \9{ \write16{ EQ \equ(#1) == #1
}}} \def \FU(#1)#2{\SIA fu,#1,#2 }
\def\etichettaa(#1){(A\veroparagrafo.\veraformula) \SIA
e,#1,(A\veroparagrafo.\veraformula) \global\advance\numfor by 1
\9{\write15{\string\FU (#1){\equ(#1)}}} \9{ \write16{ EQ \equ(#1) == #1
}}} \def\getichetta(#1){Fig.  \verafigura \SIA e,#1,{\verafigura}
\global\advance\numfig by 1 \9{\write15{\string\FU (#1){\equ(#1)}}} \9{
\write16{ Fig.  \equ(#1) ha simbolo #1 }}} \newdimen\gwidth \def\BOZZA{
\def\alato(##1){ {\vtop to \profonditastruttura{\baselineskip
\profonditastruttura\vss
\rlap{\kern-\hsize\kern-1.2truecm{$\scriptstyle##1$}}}}}
\def\galato(##1){ \gwidth=\hsize \divide\gwidth by 2 {\vtop to
\profonditastruttura{\baselineskip \profonditastruttura\vss
\rlap{\kern-\gwidth\kern-1.2truecm{$\scriptstyle##1$}}}}} }
\def\alato(#1){} \def\galato(#1){}
\def\veroparagrafo{\number\numsec}\def\veraformula{\number\numfor}
\def\verafigura{\number\numfig}
\def\geq(#1){\getichetta(#1)\galato(#1)}
\def\Eq(#1){\eqno{\etichetta(#1)\alato(#1)}}
\def\eq(#1){\etichetta(#1)\alato(#1)}
\def\Eqa(#1){\eqno{\etichettaa(#1)\alato(#1)}}
\def\eqa(#1){\etichettaa(#1)\alato(#1)}
\def\eqv(#1){\senondefinito{fu#1}$\clubsuit$#1\write16{No translation
for #1} \else\csname fu#1\endcsname\fi}
\def\equ(#1){\senondefinito{e#1}\eqv(#1)\else\csname e#1\endcsname\fi}
\openin13=#1.aux \ifeof13 \relax \else \input #1.aux \closein13\fi
\openin14=\jobname.aux \ifeof14 \relax \else \input \jobname.aux
\closein14 \fi \9{\openout15=\jobname.aux} \newskip\ttglue


\font\titolo=cmbx10 scaled \magstep1
\font\ottorm=cmr8\font\ottoi=cmmi7\font\ottosy=cmsy7
\font\ottobf=cmbx7\font\ottott=cmtt8\font\ottosl=cmsl8\font\ottoit=cmti7
\font\sixrm=cmr6\font\sixbf=cmbx7\font\sixi=cmmi7\font\sixsy=cmsy7

\font\fiverm=cmr5\font\fivesy=cmsy5\font\fivei=cmmi5\font\fivebf=cmbx5
\def\ottopunti{\def\rm{\fam0\ottorm}\textfont0=\ottorm%
\scriptfont0=\sixrm\scriptscriptfont0=\fiverm\textfont1=\ottoi%
\scriptfont1=\sixi\scriptscriptfont1=\fivei\textfont2=\ottosy%
\scriptfont2=\sixsy\scriptscriptfont2=\fivesy\textfont3=\tenex%
\scriptfont3=\tenex\scriptscriptfont3=\tenex\textfont\itfam=\ottoit%
\def\it{\fam\itfam\ottoit}\textfont\slfam=\ottosl%
\def\sl{\fam\slfam\ottosl}\textfont\ttfam=\ottott%
\def\tt{\fam\ttfam\ottott}\textfont\bffam=\ottobf%
\scriptfont\bffam=\sixbf\scriptscriptfont\bffam=\fivebf%
\def\bf{\fam\bffam\ottobf}\tt\ttglue=.5em plus.25em minus.15em%
\setbox\strutbox=\hbox{\vrule height7pt depth2pt width0pt}%
\normalbaselineskip=9pt\let\sc=\sixrm\normalbaselines\rm}
\catcode`@=11
\def\footnote#1{\edef\@sf{\spacefactor\the\spacefactor}#1\@sf
\insert\footins\bgroup\ottopunti\interlinepenalty100\let\par=\endgraf
\leftskip=0pt \rightskip=0pt \splittopskip=10pt plus 1pt minus 1pt
\floatingpenalty=20000
\smallskip\item{#1}\bgroup\strut\aftergroup\@foot\let\next}
\skip\footins=12pt plus 2pt minus 4pt\dimen\footins=30pc\catcode`@=12
\newdimen\xshift \newdimen\xwidth \newdimen\yshift
\def\ins#1#2#3{\vbox to0pt{\kern-#2 \hbox{\kern#1
#3}\vss}\nointerlineskip} \def\eqfig#1#2#3#4#5{ \par\xwidth=#1
\xshift=\hsize \advance\xshift by-\xwidth \divide\xshift by 2
\yshift=#2 \divide\yshift by 2 \line{\hglue\xshift \vbox to #2{\vfil #3
\includegraphics{#4.ps} }\hfill\raise\yshift\hbox{#5}}} \def\8{\write13}

\def\V#1{{\,\underline#1\,}}
\def\T#1{#1\kern-4pt\lower9pt\hbox{$\widetilde{}$}\kern4pt{}}
\let\dpr=\partial\def\Dpr{{\V\dpr}} \let\io=\infty\let\ig=\int
\def\fra#1#2{{#1\over#2}}\let\0=\noindent
\def\guida{\leaders\hbox to 1em{\hss.\hss}\hfill}
\def\tende#1{\vtop{\ialign{##\crcr\rightarrowfill\crcr
\noalign{\kern-1pt\nointerlineskip} \hglue3.pt${\scriptstyle
#1}$\hglue3.pt\crcr}}} \def\otto{\
{\kern-1.truept\leftarrow\kern-5.truept\to\kern-1.truept}\ }

\def\pagina{\vfill\eject}
\def\txt{\textstyle}
\def\st{\scriptscriptstyle}
\def\*{\vskip0.3truecm}

\def\lis#1{{\overline #1}}

\def\ie{\hbox{\it i.e.\ }}

\def\fiat{{}}
\def\\{\hfill\break} \def\={{ \; \equiv \; }}
\def\Re{{\rm\,Re\,}} \def\atan{{\,\rm arctg\,}}

\ifnum\aux=1\BOZZA\else\relax\fi
\ifnum\tipoformule=1\let\Eq=\eqno\def\eq{}\let\Eqa=\eqno\def\eqa{}
\def\equ{{}}\fi
\def\defi{\,{\buildrel def \over =}\,}

\def\pallino{{\0$\bullet$}}\def\1{\ifnum\mgnf=0\pagina\else\relax\fi}
\def\W#1{#1_{\kern-3pt\lower6.6truept\hbox to 1.1truemm
{$\widetilde{}$\hfill}}\kern2pt\,}
\def\Re{{\rm Re}\,}

\def\FINE{
\*
\0{\it Internet:
Author's preprints downloadable (latest version) at:

\centerline{\tt http://chimera.roma1.infn.it}
\centerline{\tt http://www.math.rutgers.edu/$\sim$giovanni}

\0Mathematical Physics Preprints (mirror) pages.\\
\sl e-mail: giovanni@ipparco.roma1.infn.it
}}

\def\NN{{\cal N}}\def\CC{{\cal C}}

\def\TT{{\cal T}}
\def\NN{{\cal N}}\def\CC{{\cal C}}

\def\HH{{\cal H}}

\def\KJ{{\bf K}}\def\dn{{\,{\rm dn}\,}}\def\sn{{\,{\rm sn}\,}}
\def\cn{{\,{\rm cn}\,}}\def\am{{\,{\rm am}\,}}\def\atan{{\,{\rm arctg}\,}}
\def\aa{{\V \a}}\def\nn{{\V\n}}\def\AA{{\V A}}

\def\oo{{\V \o}}\def\nn{{\V\n}}

\def\ndpr{{\kern1pt\raise 1pt\hbox{$\not$}\kern1pt\dpr\kern1pt}}
\def\Ndpr{{\kern1pt\raise 1pt\hbox{$\not$}\kern0.3pt\dpr\kern1pt}}
\def\palla{{$\bullet\,$}}

\fiat

\0{\titolo Hamilton--Jacobi's equation and Arnold's diffusion
near invariant tori
in a priori unstable isochronous systems.}
\*
\centerline{G. Gallavotti}
\centerline{Fisica, Roma1}
\*

\0{\bf Abstract: \it Local integrability of hyperbolic oscillators
is discussed to provide an introductory example of the Arnold's
diffusion phenomenon in a forced pendulum.}
\*
{\0\it Keywords: Hamilton Jacobi, KAM, Arnold diffusion, invariant tori}
\*
\0{\bf1. Introduction.}
\numsec=1\numfor=1\*

Here I expose, in a simple case, a well established technique. I shall
show that a Hamiltonian system composed of two clocks (\ie isochronous
rotators) and a hyperbolic oscillator can be integrated by quadratures
in a full neighborhood of the unperturbed equilibrium of the
oscillator. The rotators angular velocities form a vector
$\oo=(\o_1,\o_2)$ which we suppose Diophantine, see \S2.

The system is described by $\ell-1$ pairs $(\AA,\aa)\in
R^{\ell-1}\times T^{\ell -1}$ of ``actions'' $\AA=(A_1,\ldots,A_\ell)$
and ``angles'' $(\a_1,\ldots,\a_{\ell-1})$ and {\it one} pair of
conjugate coordinates $(p,q)$ with a Hamiltonian:

$$\oo\cdot\AA+ J(p\,q)+\e f(\aa,p,q)\Eq(1.1)$$
with $J(x)$ and $f(\aa,p,q)$ analytic for $|x|<\k$, $|p|,|q|<\k^{1/2}$
and $\aa\in T^{\ell-1}$. We suppose that $J'(0)>0$ and that $\k>0$ is
small enough so that for some $\lis g_0>0$ it is $\lis g_0<|\Re J'(x)|\le
|J'(x)|<2\lis g_0$ for all $|x|<\k$.

The Hamilton Jacobi's equation for the integration of the system is:

$$\oo\cdot\dpr_\aa\F + J((p'+\dpr_q\F)q)+
\e f(\aa,p'+\dpr_q\F,q)=\tilde J(p'(q+\dpr_{p'}\F))\Eq(1.2)$$
where the unknowns are $\F(\aa,p',q), \tilde J(x)$, and we require
that $\F,\tilde J-J$ be divisible by $\e$.

If $\F, \tilde J$ exist and are analytic in their arguments and in
$\e$, it follows that for $\e$ small enough there is a
canonical change of coordinates transforming the Hamiltonian \equ(1.1)
into:

$$\oo\cdot \AA'+\tilde J(p'q')\Eq(1.3)$$
hence the points with $p'=q'=0$ form a family of invariant tori
parameterized by $\AA'$. And the motions are completely known, and
very simple, in the new coordinates.
\*

\0{\bf Theorem 1: \it the Hamilton Jacobi's equation \equ(1.1) admits 
an analytic solution for all $\e$ small enough. The solution is
analytic in $\e$ and in $p',q$, divisible by $\e$ and so is $\tilde J-
J$.}
\*

This (form of a well known) theorem is very useful because it provides
{\it full control} of motions dwelling a very long time around the
invariant tori, in spite of the nonlinearity of the equations. This
turns out to be the key feature needed in the theory of Arnold's
diffusion.

In \S2 I prove the theorem (following [CG] where, however, it is
discussed in a much more general case). In \S3 I introduce the notion
of Arnold's diffusion, [A], in the example of a quasi--periodically
forced pendulum (\ie in a system consisting of a pendulum in
interaction with two clocks) and discuss how to apply the theorem to
show its existence. In
\S4,\S5 I provide the details, from [G3].

The example may be useful as an introduction and to clarify the
intrinsic simplicity of a problem that was solved by Arnold many years
ago and that, outside a small circle of specialists, still seems to be
(unreasonably) presented as a very hard problem.
\*
\pagina
\0{\bf\S2. A classical perturbative analysis.}
\numsec=2\numfor=1\*
Consider the Hamiltonian:

$$\HH_0=\oo\cdot\AA_0+ J_0(x_0)+f_0(\aa,p_0,q_0)\Eq(2.1)$$
with $J_0$ holomorphic in $Q_{\k_0}=\{|x_0|<\k_0\}$ and $f_0$
holomorphic in $P_{\x_0,\k_0}=\{ e^{-\x_0}<| e^{i\a_{j}}|<e^{\x_0},\,
{\rm for}\, j=1,\ldots\ell-1\}\times \{ |p_0|,|q_0|<\k_0^{1/2}\}$.
Here $\ell$ is the total number of degrees of freedom.  We assume that
$\oo$ enjoys a {\it Diophantine property}: $|\oo\cdot\nn|> C_0
|\nn|^{-\t}$ for some $C_0,\t>0$ and for all non zero integer
component vectors $\nn=(\n_1,\ldots,\n_{\ell-1})\in Z^{\ell-1}$.

We call $\e_0\defi||f_0||_{\x_0,\k_0}$ the {\it lowest upper bound} of
$|f_0|$ in its domain of definition and we call $\tilde
g_0=||J'_0||_{\k_0}$ the lowest upper bound to the derivative of $J_0$
in its domain; we also need the greatest lower bound to $\Re J_0$ that
we call $\lis g_0$. We set $\G_0=\min\{\lis g_0,C_0\}$ and by the
assumptions following \equ(1.1) it is $\lis g_0<|\Re J'(x)|\le
|J'(x)|<2\lis g_0$. Summarizing the constants $\e_0,\lis g_0,\G_0$ are
defined so that:

$$\e_0=||f_0||_{\x_0,\k_0},\qquad \lis g_0<|\Re J'(x)|\le
|J'(x)|<2\lis g_0, \qquad \G_0=\min\{\lis g_0,C_0\}\Eq(2.2)$$
The system has, therefore, two time scales, namely the pendulum time
scale $\lis g_0^{\,-1}$ and the clocks time scale $C_0^{-1}$: $\G_0$ is the
smallest.

Let us define the ``generating function'' $\AA_1\cdot\aa+ p_1
q_0+\F_0(\aa,p_1,q_0)$ via a function
$\F_0(\aa,p,q)=-(g_0(x)\ndpr+\oo\cdot\Dpr)^{-1} \tilde f_0(\aa,p,q)$
where:
\*
\0\palla we set $g_0(x)$ equal to the derivative $J'_0(x)$ of $J_0(x)$:
$g_0(x)\defi J_0'(x)$.\\
\palla we set $\ndpr\defi q\dpr_q-p\dpr_p$ and $\Dpr\defi\dpr_\aa$.\\
\palla $\tilde f_0(\aa,p,q)=f_0(\aa,p,q)-\lis f_0(pq)$ and $\lis
f_0(pq)$ is obtained from $f(\aa,p,q)$ by taking the average over
$\aa$ and then expanding in powers of $p,q$ the result and keeping
only the terms that depend on the product $pq$. This means that we
write $f(\aa,p,q)=\sum_{m,n=0}^\io f_{m,n}(\aa)p^m q^n=\sum_{m,n,\nn}
f_{\nn,m,n}e^{i\nn\cdot\aa}p^mq^n$ and set $\lis f(x)=\sum_m
f_{\V0,m,m} x^m$. In this definition we keep control of the size of
$\lis f$ because $|f_{m,n}(\aa)|< \e_0 \k_0^{-\fra12(m+n)}$ and
$|f_{\nn,m,n}|< \e_0 \k_0^{-\fra12(m+n)} e^{-\x_0|\nn|}$, where
$|\nn|=\sum_j|\n_j|$.
\*

Under the conditions on the auxiliary parameter $\d_0>0$
specified below and in the smaller domain $P_{\x_0-\d_0,\k_0
e^{-\d_0}}$, the generating map:

$$p_0= p_1+ \dpr_{q_0} \F_0(\aa,p_1,q_0),\
q_1=q_0+\dpr_{p_1}\F_0(\aa,p_1,q_0),\ \AA_0=\AA_1+\dpr_\aa
\F_0(\aa,p_1,q_0)\Eq(2.3)$$
defines a canonical map $\CC_0: P_{\x_0-\d_0,\k_0
e^{-\d_0}}\to P_{\x_0,\k_0}$:

$$\CC_0(\AA_1,p_1,q_1)=\cases{\AA_0=\AA_1+\V H_1(\aa,p_1,q_1)\cr
p_0=p_1+L_1(\aa,p_1,q_1)\cr
q_0=q_1+\tilde L_1(\aa,p_1,q_1)\cr}\Eq(2.4)$$
where $\V H_1,L_1,\tilde L_1$ are suitable functions holomorphic in
$P_{\x_0-\d_0,\k_0 e^{-\d_0}}$.

In fact by simple dimensional estimates (``Cauchy estimates''),
setting $\x'_0=\x_0-\fra12\d_0$, and $\k'_0=\k_0e^{-\fra12\d_0}$ and
choosing $B_0$ large enough it is (recall the above definition
$\e_0\defi||f_0||_{\x_0,\k_0}$; see also appendix A2):

$$\eqalign{ &||\dpr_q\F_0||_{\x'_0,\k'_0}
\ ,\ ||\dpr_{p}\F_0||_{\x'_0,\k'_0},\
\k_0^{-\fra12}||\dpr_\aa\F_0||_{\x'_0,\k'_0}
\le B_0{\G_0}^{-1}{\k_0^{-\fra12}} \d_0^{-\ell-\t-1}\e_0
\cr &||\dpr_{p}\dpr_{q}\F_0||_{\x'_0,\k'_0}
\le B_0 {\G_0}^{-1}{\k_0}^{-1} \d_0^{-\ell-\t-2}\e_0
\cr}\Eq(2.5)$$
hence under the condition $B_1 \G_0^{-1}\k_0^{-1}\e_0
\d_0^{-\ell-\t-2}<1$, with $B_1$ large enough, the map $\CC_0$ is
defined (by the implicit functions theorem) and maps
$P_{\x_0-\d_0,\k_0 e^{-\d_0}}$ into $P_{\x_0,\k_0}$.

The choice of $\F_0$ is just right so that by evaluating the
Hamiltonian in the new coordinates it takes the form:

$$\HH_1=\oo\cdot\AA_1+ J_1(p_1q_1)+ f_1(\aa,p_1,q_1)\Eq(2.6)$$
with $J_1(x)=J_0(x)+\lis f_0(x)$ and a suitable $f_1$ ``{\it which is
of order $||f_0||^2$}''.

In fact set $\x_1=\x_0-2\d_0,\k_1=\k_0 e^{-2\d_0}$ and $\lis g_1=\lis
g_0-B_2
\e_0\k_0^{-1}\d_0^{-2},\tilde g_1=2\lis g_0+B_2\e_0\k_0^{-1}\d_0^{-2}$,
with $B_2$ large (see below).

Note that $B_2\e_0\k_0^{-1}\d_0^{-2}$ is a dimensional bound for
$\fra{d}{dx}\lis f(x)$ for $|x|<\k_0 e^{-\d_0/2}$ (the $\d_0^{-2}$ is
due to the fact that the series over $p,q$ is a double series: hence
isolating the terms $(pq)^m$ and trying to bound the series expressing
$\lis f$ (via the bounds on the Fourier--Taylor coefficients
$f_{\nn,m,n}$ gives a factor $(1-e^{-\d_0/2})^{-1}$, and
furthermore one has to differentiate $\lis f$ with respect to $x$
which yields a further division by $\k_0 (1-e^{-\d_0/2})$. Then:

$$\lis g_1<|\Re J_1'|\le |J'_1|<\tilde g_1,\qquad \e_1=||f_1||_{
\x_1,\k_1}< B_3 \fra{\e_0^2}{ \G_0\k_0}\fra{\lis g_0}{\G_0}
\d_0^{-2(\ell+\t)-4}\Eq(2.7)$$
provided $B_4 \e_0\G_0^{-1}\k_0^{-1}\d_0^{-\ell-\t-1}<1$ for some
$B_4$ large enough and $\lis g_1> \fra12 \lis g_0,\tilde g_1< 4 \lis
g_0$. The bound \equ(2.7) can be checked by a first order expansion of
the Hamiltonian $\HH_0$ evaluated by expressing $p_0, \AA_1$ in terms
of $p_1,q_0,\aa$ via \equ(2.3) and by developing to first order with
respect to the increments $p_0-p_1,q_1-q_0,\AA_0-\AA_1$. The choice of
$\F_0$ is what is needed to eliminate the first order terms and the
remainders are easily estimated via Cauchy estimates, in a domain
slightly smaller than the one in which $\CC_0$ is defined, which we
recall to be $P_{\x_0-\d_0,\k_0 e^{-\d_0}}$. For details see Appendix
A2 below (see also [G1], Ch. V, \S12).

Proceeding as usual in the KAM proofs, [G1], define
$\x_j=\x_0-2\d_0-\ldots-2\d_{j-1}$ and $\k_j=\k_0
e^{-2\d_0-\ldots-2\d_{j-1}}$ and fix $\d_j=\x_0 ((j+1)4)^{-2}$. Then
it is: $\k_j\ge \k_0 e^{-\fra12\x_0}$ and $\x_j\ge\fra12\x_0$. 
We also set $\G_j=\min\{\lis g_j,C_0\}$ and $\tilde g_{j+1}=\tilde
g_j+B_2\e_j\k_j^{-1}\d_j^{-2(\ell+\t)-4}$.

In this way {\it under the conditions
$B_4\e_j\G_0^{-1}\k_0^{-1}\d_j^{-\ell-\t-1}<1$ and $2\lis g_0> \tilde
g_j\ge \lis g_j\ge \fra12 \lis g_0$} it will be: $\e_{j+1}\le B_5
{\e_j^2}{\G_0^{-1}\k_0^{-1}{\lis g_0}\G_0^{-1}\x_0^{-q}}\,
(j+1)^{q}$ having set $q\defi 2(\ell+\t)+4$.

Defining $\l_{j+1}=\l_j^2 (j+1)^{-q}$ for $j\ge1$ and $\l_0=1$ we get
$\l_j\ge e^{-q_0 2^j}$ with a $q_0$ bounded proportionally to $q$.
Furthermore defining, for a suitably large $B_5$, $\h_j=B_5\lis
g_0\G_0^{-1}\x_0^{-q}\, \l_j \e_j$ we find, from the definition of
$\l_j$:

$$ \h_j\le \h_0^{2^j}\Eq(2.8)$$
Therefore, if $B_6$ is suitably large so that the condition:

$$B_6\e_0\G_0^{-1}{\lis g_0}\G_0^{-1} \k_0^{-1}\x_0^{-q}<1,\qquad q=
2(\ell+\t)+4\Eq(2.9)$$
encompasses all the ones found so far, one can define not only
$\CC_0$ but also, recursively, $\CC_j$ in the domain $P_{\x_j,\k_j}$,
with image in $P_{\x_{j-1},\k_{j-1}}$.

The composition $\CC\defi \ldots \circ\CC_2\circ\CC_1\circ\CC_0$ of
the canonical maps will be well defined and it will give as a result a
map $\CC:\,(\AA_0,p_0,q_0)\otto (\AA_\io,p_\io,q_\io)$ still of the
form \equ(2.4) with some function $\V H,L,\tilde L$ analytic in
$P_{\fra12\x_0,\k_0 e^{-\fra12\x_0}}$ and transforming the Hamiltonian
\equ(2.1) into the ``normal form'':

$$\HH_\io=\oo\cdot\AA_\io+ J_\io(p_\io q_\io)\Eq(2.10)$$
with $J_\io$ holomorphic in $Q_{\k_0 e^{-\fra12\x_0}}$ wit $\fra12 \lis g_0<
|\Re J'_\io|\le |J'_\io|<4\lis g_0$.

{\it Comments:} if $f_0(\aa,p,q)=f(-\aa,q,p)$ then the symmetries of
the problem imply that:

$$\V H_\io(\aa,p,q)=\V H_\io(-\aa,q,p), \ L_\io(\aa,p,q)=\tilde
L_\io(-\aa,q,p)\Eq(2.11)$$
a limit as $j\to\io$ of the corresponding relations which also hold
identically for $\V H_j,L_j,\tilde L_j$, for
$j=1,2,\ldots$. Furthermore $\V H$ has by construction $\V0$ average
when evaluated at $p=q=0$ (a property of ``twistless tori'', see [G2])
or, more generally, at $pq=0$. I shall call, therefore, $\AA_\io$ the
{\it average actions} of such torus.
\*

\0{\bf \S3. An application.}
\numsec=3\numfor=1\*

We consider a system with Hamiltonian:

$$\HH=\oo\cdot\AA+\fra{I^2}{2}+g^2(\cos\f-1)+\e f(\f,\aa)\Eq(3.1)$$
where $(I,\f)\in R\times T^1$ describe a pendulum while $(\AA,\aa)\in
R^2\times T^2$ describe two clocks and $f$ is an interaction which we
take to be a even trigonometric polynomial in $\aa=(\a_1,\a_2),\f$. A
non trivial case is $f=\cos(\a_1+\f)+\cos(\a_2+\f)$.

Near the pendulum equilibrium one can use Jacobi's coordinates $p,q$,
see Appendix A1. In such {\it local coordinates} the system has
Hamiltonian:

$$\HH'=\oo\cdot\AA+ J(pq)+\e f(\aa,p,q)\Eq(3.2)$$

Hence by the theorem above we can find a (local) system of coordinates
$\AA_\io,\aa,p_\io,q_\io$ in which the motion is described by a trivial
Hamiltonian \equ(2.10). This means that it is:

$$\AA_\io=const,\ \aa\to\aa+\oo t, \ p_\io\to p_\io e^{-\tilde g(x)t},
\ q_\io\to q_\io e^{\tilde g(x)t}\Eq(3.3)$$
which shows that the set $p_\io=q_\io=0$ is an invariant torus
parameterized by $\AA_\io$ and denoted $\TT(\AA_\io)$.

Physically this is a very special set of motions in which the pendulum
{\it does not ever fall down from its unstable equilibrium position}
but performs small oscillations around it. It also shows that the sets
$q=0$, denoted $W^s(\AA_\io)$, or $p=0$, denoted $W^u(\AA_\io)$, are
invariant manifolds for $t>0$ and $t<0$ respectively, such that data
on them evolve towards the (quasi periodic) motions on the invariant
torus $\TT(\AA_\io)$, respectively, as $t\to\pm\io$.

The meaning of $\oo\cdot\AA$ is the ``energy of the springs that move
the clocks''. {\it The problem of Arnold's diffusion is whether one can
find motions whose net effect is to transfer energy from one reservoir
to the other for all $\e>0$ no matter how small}: note that if $\e=0$
transfer is impossible.

Mathematically this is a motion that starts near a quasi periodic
motion with average action $\AA=\AA_0$ and ends up, after a finite
time, near one with average action $\AA_1\ne\AA_0$. The key point in
this definition is that Arnold's diffusion between $\AA_0$ and $\AA_1$
if such motions exist {\it for all $\e>0$ small enough}, no matter how
small (provided $\e\ne0$ and, of course, for suitable $\e$--dependent
initial data). Here ``near'' means closer than the half the distance
between the tori corresponding to $\AA_0$ and $\AA_1$.

Of course for this to be possible it must be that the extreme motions
have the same energy: it is a simple {\it symmetry property} of the
above models (with $f$ even) that the condition that two quasi
periodic motions with parameters $\AA_\io=\AA_0$ and $\AA_\io=\AA_1$
have the same energy is, generically and if $\e$ is small, simply
$\oo\cdot(\AA_1-\AA_0)=0$, see [G3].

Consider the line $s\to\AA(s)=\AA_0+(\AA_1-\AA_0)s$, $s\in [0,1]$,
supposing it orthogonal to $\oo$.  One says that there is a
heteroclinic chain between $\AA_0$ and $\AA_1$ if there are $\NN$
values $s_0=0,s_1,\ldots s_\NN$ such that for all $j$'s there exists a
``heteroclinic'' motion, \ie a motion that is asymptotic as $t\to-\io$
to the quasi periodic motion on the torus with average actions
$\AA_\io=\AA(s_j)$ and that is asymptotic as $t\to+\io$ to the quasi
periodic motion on the torus with average actions
$\AA_\io=\AA(s_{j+1})$.

It is a simple application of the implicit functions theorem to show
that any chain with $s_{j+1}-s_j$ small enough is a heteroclinic chain,
see [G3], for generic $f$. In fact if $f=\cos(\a_1+\f)+\cos(\a_2+\f)$
this is the case, see [CG], [GGM]. The number $\NN$ will be in this
case, and in general, of size $O(\e^{-1})$, see [GGM].

How small $\e$ has to be so that it is ``small enough''? this depends
on $g$. An interesting question is how small has $\e$ to be compared
to $g$ (the pendulum characteristic frequency).

Consider a simple case: $\oo=(\h,1)$ and $g^2=\h$ and $f(\aa,\f)=
\b\cos(\a_2+\f)+\e \h^c \cos(\a_1+\f)$. In this case if $c$ is large
enough taking $\b=\e \h^c$ is sufficient, see [GGM]. However one can
do much better and allow $\b$ to be larger than $1$, {\it\ie
independent of $\e$}! provided one excludes {\it finitely many}
possibly exceptional values of $\b$, see [GGM]. This shows that a
often stated condition that $\b$ has to be exponentially small in
$\h^{-1/2}$ can be improved in three time scales problems.

We now study the existence of diffusion reproducing for convenience
of the reader \S4 and \S5 of [G3].
\*

\0{\bf\S4. Geometric concepts.}
\numsec=4\numfor=1\*

Let $2\,\k>0$ be smaller than the radius of the disk in the $(p,q)$
plane where the canonical coordinates $(\AA_\io,\aa,p_\io,q_\io)$ can
be used. I will drop soon the subscripts $\io$ to simplify the
notation. To visualize the geometry of the problem we shall need the
following geometric objects:

\*
\pallino(a) a point $X_i$, heteroclinic between the torus
parameterized by $\AA_\io=\AA(s_i)\defi\AA_i$ which we shall denote
$\TT(\AA_i)$ and the ``next'' torus $\TT(\AA_{i+1})$: this is a point
$X_i$ such that $S_t X_i\tende{t\to+\io}$ $
\TT(\AA_{i+1})$ and $S_t X_i\tende{t\to-\io}
\TT(\AA_{i})$. We denote the local coordinates
of $X_i$ as  $X_i=(\AA_i,\aa_i,0,\k)$.
\*

\pallino(b) we can extend the manifolds $W^s(\TT(\AA_i))$ and
$W^u(\TT(\AA_i))$ by applying to them the time evolution flow $S_t$.
Then by our assumption the $\AA_i$ form a heteroclinic chain and
therefore $W^s(\TT(\AA_{i+1}))$ evolves in a finite (negative) time to
the vicinity of $\TT(\AA_{i})$, and in fact its image (at a properly
chosen negative time) will contain $X_i$. Hence it can be described in
the local coordinates near such torus. The equations, at fixed $q=\k$,
of the connected part of $W^s(\AA_{i+1})$ containing $X_i$, will be
written in the local coordinates near $\TT(\AA_i)$ as:

$$Y_i(\aa)=(\AA^s_{i+1}(\aa),\aa, p^s_{i+1}(\aa),\k)\Eq(4.1)$$
with $|\aa-\aa_i|<\z$ for some $\z>0$ ($i$--independent): it is
$\AA^s_{i+1}(\aa_i)=\AA_i, \, p^s_{i+1}(\aa_i)=0$ because we require
$Y_i(\aa_i)=X_i$. There are constants $F',F$ such that
$|\AA^s_{i+1}(\aa)-\AA^s_{i+1}(\aa_i)|$ and
$\max_{|\aa-\aa_i|=\,fixed}|p^s_{i+1}(\aa)|$ are bounded, for $\z$
small enough, below by $F'|\aa-\aa_i|$ and above by $F |\aa-\aa_i|$;
the constants $F',F$ are generically positive, see [GGM] for
instance. They are positive in the (non trivial) example
$f=\cos(\a_1+\f) +\cos(\a_2+\f)$.

Note that $W^s(\AA_{i+1})$ also contains a part with local equations
$(\AA_{i+1},\aa,p,0)$ which is {\it not} to be confused with the
previous one described by the function $Y_i(\aa)$. This is more easily
understood by looking at the meaning of the above objects in the
original $(\AA,\aa,I,\f)$ coordinates: in a way the first part of
$W^s(\AA_{i+1})$ is close to $\f=0$ and the second to $\f=2\p$. They
can be close to each other because of the periodicity of $\f$, but
they are conceptually quite different.

\*
\pallino(c) a point $P_i=Y_i(\tilde\aa_i)$ with $|\tilde\aa_i -\aa_i|=r_i$,
where $\tilde\aa_i,r_i$ will be determined recursively, and a
neighborhood $B_i$:
\kern-5pt
$$B_i=\{|\AA-\AA^s_{i+1}(\aa)|<\r_i,\
|\aa-\tilde\aa_i|<\r_i,
\ |p^s_{i+1}(\aa)-p|<\r_i,\ q=\k\}\Eq(4.2)$$
\nobreak
where $\r_i<r_i$ is another length to be determined recursively. If
$\lis g\defi\fra12\lis g_0, 4 \lis g_0$ are a lower and upper bound to
$\tilde J'(x)$ for $|x|<4\k^2$, the point $P_i$ evolves in a time
$T_i\simeq
\lis g^{\,-1} \log \k^{-1}$ into a point $X'_i$ near $\TT(\AA'_{i+1})$
which has local coordinates $X'_i=(\AA_{i+1},\aa'_i,\k,0)$.
\*

\pallino(d) The points $\x$ of the set $B_i$ are mapped by the
time evolution to points that, at the beginning at least, come close
to $\TT(\AA_{i+1})$ and in a time $\t(\x)$ acquire local coordinates
near $\TT(\AA_{i+1})$ with $p=\k$ exactly: the time $\t(\x)$ is of the
order of $\lis g^{\,-1}\log\k^{-1}$.

If $S_t$ is the time evolution flow for the system \equ(1.1) we write
$S\x=S_{\t(\x)}\x$ (note that $S$ depends also on $i$). Then $S$ maps
the set $B_i$ into a set $SB_i$ containing:

$$B'_i=\{|\AA-\AA_{i+1}|<\fra1E \r_i,\ |\aa-\aa'_i|<\fra1E \r_i,
\ p=\k,\ |q|<\fra1E \r_i\}\Eq(4.3)$$
because all the points in $B_i$ with $\AA=\AA^s_{i+1}(\aa),\, p=
p^s_{i+1}(\aa), \, q=\k$ evolve to points with $\AA=\AA_{i+1}$,
$p=\k$, $q=0$ and $\aa$ close to $\aa'_i$, by the definitions. Here
$E$ is a bound on the jacobian matrix of $S$ (which, being essentially
a flow over a time $O(\lis g^{\,-1}\log\k^{-1})$, has derivatives bounded
$i$--independently: since we suppose that $\e$ is ``small enough'' we
could take $E=1+b\e$ for some $b>0$ if $|\AA_i-\AA_{i+1}|<O(\e)$).
\*

\0{\bf\S5. The [CG]-method of proof of the theorem.}
\numsec=5\numfor=1\*

Let $\NN$ be the number of elements of the heteroclinic chain.
Consider the points $Y^s_{i+1}(\aa)\in W^s(\AA_{i+2})$ with
coordinates $(\AA^s_{i+2}(\aa),\aa,p^s_{i+2}(\aa),\k)$. They will
evolve backwards in time so that $\AA$ stays constant, $\aa$ evolves
quasi-periodically hence ``rigidly'', and $p^s_{i+2}(\aa)$ evolves to
$\k$ while the $q$--coordinate evolves from $\k$ to $q=p^s_{i+2}(\aa)$
(because $pq$ stays constant, see \equ(2.1)). This requires
going backward in time by an amount of the order of $T_\aa\simeq \lis
g^{\,-1}\log \k|p^s_{i+2}(\aa)|^{-1}\tende{\aa\to\aa_{i+1}}+\io$.

Therefore there is a sequence $\aa^n\ne \aa_{i+1}$ such that $\aa^n\to
\aa_{i+1}$, $p^s_{i+2}(\aa^n)\to0$ $\AA^s_{i+2}(\aa^n)\to
\AA_{i+1}$ {\it and} $\aa^n-\oo T_{\aa^n}\tende{n\to\io}\aa'_i$, as
a consequence of the Diophantine properties of $\oo$. So that there is
$\tilde\aa_{i+1}\defi \aa^n$ with $n$ large enough and a
point $P_{i+1}=(\AA^s_{i+2}(\tilde\aa_{i+1}),\tilde\aa_{i+1},
p^s_{i+2}(\tilde\aa_{i+1}),\k)\in W^s(\AA_{i+2})$ (actually
infinitely many) which evol\-ves, backwards in time, from $P_{i+1}$ to a
point of $B'_i$.

Hence we can define $r_{i+1}=|\tilde\aa_{i+1}-\aa_{i+1}|$ and
$\r_{i+1}$ small enough so that the backward motion of the points in
$B_{i+1}$ enters in due time into $B'_i$.  It follows that the set
$B_i$ evolves in time so that all the points of $B_{i+1}$ are on
trajectories of points of $B_i$. Hence all points of $B_{\NN}$ will be
reached by points starting in $B_0$.

This completes the proof. All constants can be computed explicitly,
even though this is somewhat long and cumbersome.  The
result is an extremely large diffusion time $T$ (namely $\lis g^{-1}$ times
the value at $\NN$ of a composition of $\NN$ exponentials! at least
this is the estimate I get after correcting an error in \S8 of [CG]:
the error was minor but its simple correction leads to substantially
worse bounds, see [CG]).

\*
\0{\bf\S6. Concluding remarks:}
\numsec=6\numfor=1\*

\0(1) The above proof of Arnold's diffusion is very simple but it
leads to a ``bad'' bound. One can improve the bound by using that
there are many invariant tori and one can choose around which to
construct a diffusing trajectory (for $\e$ small enough). The bound
goes down to $g^{-1}O(2^\NN)$, see [G3].

\0(2) This is still far from the best bounds in the literature for similar
problems, [Be], [Br], obtained by different methods.

\0(3) The above (obvious) proof is likely to be what Arnold, [A], had in
mind when he proposed his example: the example has in common with the
above \equ(1.1) {\it the key feature} of having a coordinate system
like the one discussed in \S2. The corresponding local Hamilton
Jacobi's equation admits a solution ``with no exceptions'', one says a
``gap-less system of coordinates'', trivializing the flow: usually
when the system is not isochronous or it is not of the Arnold's type
there are open regions of phase space where the nice coordinates in
which the flow is trivial cannot be defined, see [CG] for what can be
done in such cases.

\0(4) The role of the special coordinates $(\AA_\io,\aa,p_\io,q_\io)$
defined after \equ(2.9) is {\it essential}: it is clear from the
analysis of \S5 that one must control the motions near the tori over a
very long time span. Since the equations of motion in the original
coordinates are non linear this would be impossible in absence of an
{\it exact and simple} solution, as the one given by \equ(3.3): any
small perturbation of the initial data would be amplified
exponentially in time in an uncontrollable way. The \equ(3.3),
instead, confine the expansion to either the $p$ or $q$ coordinate and
in a very explicit way. Hence the above is a good example of a non
trivial use of the Hamilton--Jacobi's equation.

\0(5) The method discussed in \S5 is a simplified version of a method
known as ``windowing'' described in a early work, [Ea], and developed
in more recent works [M], [C]. But it is much less ambitious and
detailed and the bounds obtained in the last two papers are far better
than the ones we achieve here. I am indebted to P. Lochak and
J. Cresson for pointing out the relevance of the latter papers.

\*
\0{\bf Acknowledgments: \it This is a contribution to the ISI school
on dynamical systems of June 1997 in Torino: I am indebted to
G. Gentile and V. Mastropietro for many discussions and suggestions. I
thank A. Ambrosetti and E. Vesentini for the invitation to lecture at
ISI.}
\*

\0{\bf Appendix A1. Jacobi's map.}
\numsec=1\numfor=1\*

This appendix is standard: here it is taken from A9 of [CG] with small
changes, to use it for future references.

The theory of Jacobian elliptic functions shows how to perform a
complete calculation of the functions, below denoted $R,S$, in terms
of which the canonical Jacobi's coordinates are defined, see [GR]
(9.198),(9.153), (9.146), (9.128), (9.197).  The result, reported
for completeness, is discussed in terms of the pendulum
energy:
$${\dot\f^2\over2}+g^2(1-\cos\f)=E\Eqa(A1.1)$$
where the origin in $\f$ is set at the stable equilibrium, to adhere to
the notations in the theory of elliptic functions.
Setting $u=t(E/2)^{1/2}\=\e^{1/2} gt$, $k^2={2 g^2/E}=\e^{-1}$
where $\e$ is the {\it dimensionless} energy so that $\e=1$ is the
separatrix, let:
$$\KJ(k)=\ig_0^{\p/2} {d\a\over (1-k^2\sin^2\a)^{1/2}}\Eqa(A1.2)$$
We shall use the ``standard'' notations (\ie those in
[GR]) for the Jacobian elliptic integrals {\it except} for $x(.)$, which is
usually denoted $q(.)$, but which we would confuse with the
variable $q$ that we want to construct:
$$\eqalign{ k'=&(1-k^2)^{1/2},\qquad g_J=g{\p\over2k\KJ(k')}=\e^{1/2}
g,\qquad \l\={1\over2}{1-k^{1/2}\over1+k^{1/2}}\cr
x(k')=&e^{-\p\KJ(k)/\KJ(k')}= \l+2\l^5+15\l^9+150\l^{13}+1707\l^{17}+
\ldots\cr}\Eqa(A1.3)$$

In terms of the above notations we have, directly from the definitions
(\ie from the equations of motion):
$$
I(t)=\dot \f=-2g\e^{1/2}\dn(u,k),\qquad
\f(t)=2\am(tg\e^{1/2})\Eqa(A1.4)$$
which yield, changing the origin for $\f$ to the
unstable point to conform with our notations (\ie obtaining
$\f(t)=2(\am(tg\e^{1/2})+\p/2)$), for $I(t)=R(p(t),q(t)),
\f(t)=S(p(t),q(t))$:
$$R=-2g\e^{1/2}{\dn(iu,k')\over\cn(iu,k')},\quad
\sin {S\over2}={1\over \cn(iu,k')},\quad\cos {S\over2}=
i{\sn(iu,k')\over\cn(iu,k')}
\Eqa(A1.5)$$

Setting $p=e^{-g_Jt}, q=x(k') e^{g_J t}$, see [GR], and using
$R(p,q)=g_J (-p\dpr_p+q\dpr_q)S(p,q)$ to evaluate $S$ from $R$, the
quoted basic relations between elliptic integrals imply
immediately that the $I(t)=R(p(t),q(t)), \f(t)=S(p(t),q(t))$,
solve the pendulum equations if:
$$\eqalignno{
\txt R(p,q)=&\txt -2g_J
\Bigl[{p\over1+p^2}+{q\over 1+q^2}-\sum_{n=1}^\io(-1)^n{1+x^{2n-1}\over
1-x^{2n-1}}(p^{2n-1}+q^{2n-1})\Bigr]\cr
\txt S(p,q)=&\txt 2\left[\atan p-\atan q-\sum_{n=1}^\io(-1)^n{1+x^{2n-1}\over
1-x^{2n-1}}{(p^{2n-1}-q^{2n-1})\over 2n-1}\right]&\eqa(A1.6)\cr
\txt \sin {S(p,q)\over2}=&\txt \fra{g_J}{g}
\Bigl[
{p\over 1+p^2}-{q\over 1+q^2}-\sum_{n=1}^\io(-1)^n{1-x^{2n-1}\over
1+x^{2n-1}}(p^{2n-1}-q^{2n-1})\Bigr]\cr
\txt \cos {S(p,q)\over2}=&\txt -\fra{g_J}{2g}
\Bigl[
{1-p^2\over 1+p^2}+{1-q^2\over 1+q^2}+2
\sum_{n=1}^\io(-1)^n{1-x^{2n}\over
1+x^{2n}}(p^{2n}+q^{2n})\Bigr]\cr}$$
with $x\=pq$.  Note that $g_J$ depends on $x$, and so do $k',k$: hence
the coefficients of the first and of the last two of \equ(A1.6) are
also functions of $x=pq$.  Furthermore the (dimensionless) energy $\e$
becomes a function $\e(\x)$ of $\x=pq$ defined by inverting the map:
$$\x=x(k')\=x((1-\e^{-1})^{1/2})\Eqa(A1.7)$$
and the point corresponding to $\f=\p$ and to a
dimensionless energy $\e$, has coordinates:
$$p\=1,\quad q\=x(k')\Eqa(A1.8)$$
(a rearrangement of \equ(A1.6) showing convergence for $p=1$ and
$|x|<1$ is exhibited below).

The variables $(p,q)$ defined above are nice and natural: however they
are not canonically conjugated to $(I,\f)$: the Jacobian determinant
of the map $(p,q)\otto(I,\f)$ is not $1$. But the Jacobian determinant
must be a function $D(x)=\fra{\dpr (p,q)}{\dpr(I,\f)}$ of $x$ alone
(\ie of the product $pq$); then \equ(A1.8) and the equations of motion
imply that $D(x)^{-1}= g_J^{-1}\fra{2g^2d\e(x)}{dx}=4g
\fra{d\e^{1/2}}{dx}$.

Therefore one can modify the variables $p,q$ into new variables
$(p_J,q_J)=(p F(x), q F(x))$ with $F$ such that the Jacobian
$\fra{\dpr (p_J,q_J)}{\dpr(I,\f)}=\fra{\dpr (p_J,q_J)}{\dpr(p,q)}
D(x)$, which is $D(x)\cdot\dpr_x(x F^2(x))$, is identically $1$. One
finds: $F(x)=(4g)^{1/2}(\fra{\e(x)^{1/2}-1}x)^{1/2}$.

To invert the map $(p_J,q_J)=(p F(x), q F(x))$ define $x_J\defi
p_Jq_J$ and $G(x_J)\defi F(x)^{-1}$ then: $p=p_J G(x_J)$ and $q=q_J
G(x_J)$, $x=x_J G^2(x_J)$.  The final result is a local canonical map
between Jacobi's coordinates $(p_J,q_J)$ and global $(I,\f)$
coordinates:
$$I=R(p_JG(x_J),q_JG(x_J)),\qquad
\f=S(p_JG(x_J),q_JG(x_J))\Eqa(A1.9)$$
where $R,S$ are defined above, see \equ(A1.6) which are written in a
form easily recognized in the elliptic functions tables.  The
functions $R,S$ can be rewritten in the following form:
\def\txt{\textstyle}
$$\eqalign{\txt
R(p,q)=&\txt-4g\left[\sum_{m=0}^\io\bigr({x^mp\over 1+x^{2m}p^2}
+{x^mq\over 1+x^{2m}q^2}\bigl)\right]\cr\txt
S(p,q)=&\txt4\left[\sum_{m=0}^\io\bigl(\atan x^mp-\atan
x^mq\bigr)\right]\cr
\txt
\sin {S(p,q)\over2}=&\txt
\fra{2g_J(x)}{g}                         
\left[\sum_{m=0}^\io (-1)^m\bigl(
{x^m p\over 1+ x^{2m} p^2}-{x^m q\over 1+x^{2m}q^2}\bigr)\right]\cr\txt
\cos {S(p,q)\over2}=&\txt
\fra{g_J(x)}{2g}                         
\left[1-2\sum_{m=0}^{\io}(-1)^m\bigl({x^{2m}p^2\over 1+x^{2m}p^2}+
{x^{2m}q^2\over 1+x^{2m}q^2}\bigr)\right]\cr}\Eqa(A1.10)$$
exhibiting some of the properties of the Jacobi's map in a better way.

One checks that in the $(p_J,q_J)$ variables the pendulum Hamiltonian,
\equ(A1.1) has become a function $J(p_Jq_J)=2g^2+ g x_J+O(x_J^2)$. The
domain of definition of the map is given by the properties of the
elliptic functions or, more restrictively, by the domain of
convergence of the above series. It includes a disk of some radius
$\r_J>0$ around the origin.

The important symmetry $R(p,q)=R(q,p)$ and $S(p,q)=-S(q,p)$ is
manifest.
\pagina
\*
\0{\bf Appendix A2: Dimensional bounds.}
\numsec=2\numfor=1\*

We give here two examples of dimensional estimates.

\palla The bounds \equ(2.5) are obtained by writing (with $\nn\in Z^{\ell-1}$):

$$\F_0(\aa,p_1,q_0)=-\sum_{\nn,n,m\atop |\nn|+|n-m|>0} f_{\nn,n,m}
\fra{e^{i\nn\cdot\aa}p_1^mq_0^n}{i\oo\cdot\nn+g(p_1q_0)(n-m)}\Eqa(A21.1)$$
so that setting $\x^{\prime\prime}_0=\x_0-\fra{\d_0}4,\k^{\prime\prime}_0=
\k_-e^{-\fra{\d_0}4}$ and $\x'_0=\x_0-\fra{\d_0}2,
\k^{\prime}_0=\k_0e^{-\fra{\d_0}2}$ (assuming $\ell\ge3$ and
using the bound on $|f_{\nn,m,n}|$ preceding \equ(2.3)): 

$$\eqalign{
||\F_0||_{\x^{\prime\prime}_0,\k^{\prime\prime}_0}\le& \sum_{\nn,m,n}
\Big(\fra{|\nn|^\t\d_{m=n}}{C_0}+\fra{\d_{m\ne n}}{\lis g_0}\Big)
e^{-\fra14\d_0 |\nn|}e^{-\fra14\d_0 (n+m)}\,\e_0\le
B_1\fra{\e_0}{\G_0\d_0^{\t+\ell}}\cr}\Eqa(A2.2)$$
for a suitably chosen $B_1$; hence \equ(2.5) follows because, for
instance, $|\dpr_q
\F_0||_{\x'_0,\k'_0}$ can be bounded above by $const\,{\k_0^{-\fra12}}
||\F_0||_{\x^{\prime\prime}_0,\k^{\prime\prime}_0}$ (a typical example
of a dimensional estimate).

\palla By using \equ(2.3) one immediately finds:
$$\eqalign{
&\HH_0=\HH_1+\big(\oo\cdot\Dpr \F_0 + g_0(p_1q_0)\Ndpr\F_0\big)+
f_0(\aa,p_1,q_0)-\lis f_0(p_1\,q_0)+\cr
&+\big[J_0((p_1+\dpr_{q_0}\F_0)q_0)-J_0(p_1\,q_0)-
J'_0(p_1\,q_0)\, q_0\,\dpr_{q_0}\F_0\big]+\cr
&-\big[J_0(p_1(q_0-\dpr_{p_1}\F_0))-J_0(p_1\,q_0)+
J'_0(p_1\,q_0)\,p_1\,\dpr_{p_1}\F_0\big]+\cr&+
\big\{f_0(\aa,p_1+\dpr_{q_0}\F_0,q_0)-f_0(\aa,p_1,q_0)\big\}-
\big\{\lis f_0(p_1\,( q_0+\dpr_{p_1}\F_0))-\lis f_0(p_1\,q_0)\big\}\cr}
\Eqa(A2.3)$$

The terms in square and curly brackets can be bounded dimensionally:
for instance the first term in square brackets is bounded in the domain
$P_{\x_0-2\d_0,\k_0e^{-2\d_0}}$ by a bound on the second
derivative of $J_0$ in $P_{\x_0-\d_0,\k_0 e^{-\d_0}}$ (\ie
$2\,\lis g_0 $ times
$\k_0^{-1}(1-e^{-\d_0})^{-1}$ times a bound (on $P_{\x_0-2\d_0, \k_0
e^{-2\d_0}}$) of the square
of $p_1\dpr_{p_1}\F_0$ or $q_0\dpr_{q_0}\F_0$ that is given by
\equ(2.5)': provided $|p_1\dpr_{p_1}\F_0|, |q_0\dpr_{q_0}\F_0|< \d_0$
in $P_{\x_0-2\d_0,\k_0 e^{-2\d_0}}$ which gives the condition following
\equ(2.7)).

Likewise the terms in curly brackets can be bounded. The final result
is \equ(2.7).
\*
\*

\0{\it References.}
\*

\item{[A] } Arnold, V.: {\it Instability of dynamical systems with several
degrees of freedom}, Sov. Mathematical Dokl., 5, 581-585, 1966.

\item{[Be] } Bessi, U.: {it An approach to Arnold's diffusion through
the Calculus of Variations}, Nonlinear Analysis, 1995.

\item{[Br] } Bernard, P.: {\it Perturbation d'un hamiltonien
partiellement hyperbolique}, C.R. Academie des Sciences de Paris,
{\bf 323}, I, 189--194, 1996.

\item{[C] } Cresson, J.: {\it Symbolic dynamics for homoclinic
partially hyperbolic tori and ``Arnold diffusion''}, Preprint of
Institut de mathematiques de Jussieux, 1997.

\item{[CG] } Chierchia, L., Gallavotti, G.: {\it Drift and diffusion in
phase space}, Annales de l' Institut Poincar\`e, B, {\bf 60}, 1--144,
1994.

\item{[Ea] } Easton, R.W.: {\it Orbit structurenear trajectories
biasymptotic to invariant tori}, in {\sl Classical mechanics and
dynamical systems}, ed. R. Devaney, Z. Nitecki, p. 55--67, Dekker,
1981.

\item{[G1] } Gallavotti, G.: {The elements of mechanics}, Springer, 1983.

\item{[G2] } Gallavotti, G.: {\it Twistless KAM tori}, Communications in
Mathematical Physics {\bf 164}, 145--156, (1994).

\item{[G3] } Gallavotti, G.: {\it Fast Arnold's diffusion in
isochronous systems}, preprint, mp$\_$arc@math. utexas.edu  \# 97-481,
and chao-dyn@xyz.lanl.gov \# 9709011.

\item{[M] } Marco, J.P.: {\it Transitions le long des chaines de tores
invariants pour les syst\`emes hamiltoniens analytiques}, Annales de
l' Institut Poincar\'e, {\bf64}, 205--252, 1995.
\FINE
\*
\0Archived also in:\\
mp$\_$arc@math.utexas.edu        \# 97-???    and\\
chao-dyn@xyz.lanl.gov \kern0.4cm \# 970????
\end